\batchmode
\makeatletter
\def\input@path{{/home/subhro/PHYSICALREVE/Resubmission//}}
\makeatother
\documentclass[11pt]{article}
\usepackage{times}
\usepackage[T1]{fontenc}
\usepackage{geometry}
\usepackage{graphics}
\usepackage{setspace}
\makeatletter

\providecommand{\LyX}{L\kern-.1667em\lower.25em\hbox{Y}\kern-.125emX\@}
\newcommand{\noun}[1]{\textsc{#1}}

 \newcommand{\lyxaddress}[1]{
   \par {\raggedright #1 
   \vspace{1.4em}
   \noindent\par}
 }

\makeatother

\begin{document}

\title{Phase Transitions In Two Planar Lattice Models And Topological Defects: A Monte
Carlo Study}

\author{Subhrajit Dutta and Soumen Kumar Roy}

\maketitle

\lyxaddress{\centering Department of Physics, Jadavpur University, Calcutta\( - \)700
032, India}

\begin{abstract}
{\large Monte Carlo simulation has been performed in the planar P\( _{2} \)
and P\( _{4} \) models to investigate the effects of the suppression of topological
defects on the phase transition exhibited by these models. Suppression of the
1/2-defects on the square plaquettes in the P\( _{2} \) model leads to complete
elimination of the phase transition observed in this model. However in the P\( _{4} \)
model, on suppressing the single 1/2-defects on square plaquettes, the otherwise
first order phase transition changes to a second order one which occurs at a
higher temperature and this is due to presence of large number of 1/2-pair defects
which are left within the square plaquettes. When we suppressed these charges
too, complete elimination of phase transition was observed.}{\large \par}
\end{abstract}

\section{{\large INTRODUCTION}\large }

{\large It is well known that conventional long range order cannot exist in
a two dimensional continuous spin system\cite{mermin}. However the existence
of topological charges leads to a quasi long range order (QLRO) to disorder
phase transition. The most notable and thoroughly investigated example is the
two dimensional XY model (O(2) model) in which, using a renormalization group
technique, Kosterlitz and Thouless \cite{kost} predicted a QLRO-disorder phase
transition which is associated with the unbinding of the vortex-antivortex pairs
(topological charges of strength \( \pm  \)1) which are stable topological
defects in this system. The phase with QLRO is characterized by an algebraic
decay of the spin-spin correlation function which is a slower decay than the
fast exponential one which is observed in a completely disordered system. On
the other hand in the two dimensional nonlinear sigma model, an example of which
is the planar O(3) model,  there exists no stable topological defect and the
system remains disordered at all finite temperature \cite{2dO(3)}.}
\begin{table}
{\centering \begin{tabular}{|c|c|c|c|}
\hline 
T&
The charges suppressed&
p&
b\\
\hline 
\hline 
0. 48&
single 1/2 -suppressed on square plaquettes&
0. 688&
0. 275\\
\hline 
0. 48&
total&
0. 432&
0. 385\\
\hline 
0. 50&
single 1/2 -suppressed on square plaquettes&
0. 615&
0. 062\\
\hline 
0. 50&
total&
0. 474&
0. 353\\
\hline 
0. 68&
single 1/2 -suppressed on square plaquettes&
1. 410&
0\\
\hline 
0. 68&
total &
0. 650&
0\\
\hline 
\end{tabular}\par}

\caption{The parameters obtained in the L=80, P\protect\( _{4}\protect \) model for
the best fit g(r)\protect\( =ar^{-p}+b\protect \) of the correlation function
at the temperatures indicated. In the normal lattice at these temperatures g(r)
decays exponentially to zero. }
\end{table}
{\large \par}

{\large Another class of the two dimensional systems of interest is the planar
Lebwohl Lasher (LL) \cite{lebwohl} model and a modified version of it to be
elaborated below. In the 3-dimensional version of the LL model, the spins (of
dimensionality 3), located at the sites of a simple cubic lattice, interact
with nearest neighbors via a potential \( - \)\( P_{2}(cos\theta ) \) where
\( \theta  \) is the angle between the spins and P\( _{2} \) is the second
Legendre polynomial. This model successfully describes the orientational aspects
of a nematic and undergoes a weakly first order phase transition, representative
of the nematic\( - \)isotropic transition, seen in a real nematic. A number
of investigators \cite{zhang,pelco} have used a modified version of the LL
model by adding a \( - \)P\( _{4} \) term to the usual \( - \)P\( _{2} \)
one, P\( _{4} \) being the fourth order Legendre polynomial. The introduction
of the P\( _{4} \) term reduces the sharpness of the peak of the P\( _{2} \)
term in the potential at \( \theta =\pi /2 \) and may lead to the appearance
of a local minimum, depending on the relative strengths of the P\( _{2} \)
and P\( _{4} \) terms in the potential. This is found to enhance the first
orderdness of the N-I transition. }{\large \par}

{\large The two dimensional version of the LL model and a modified version of
it with a pure P\( _{4} \) interaction between nearest neighbor spins have
recently been investigated using Monte Carlo(MC) methods by a number of authors
\cite{kunz,enakshi,abhijit}. In the rest of this paper we shall refer to these
as the planar P\( _{2} \) and P\( _{4} \) models respectively. Both models
possess in addition to the usual O(3) symmetry, a Z\( _{2} \) symmetry as well
and this leads to the identification of the antipodal points in the order parameter
space S\( ^{2} \). The planar P\( _{2} \) model is known to exhibit a continuous
phase transition at a dimensionless temperature whose thermodynamic limit is
0.547 \cite{enakshi} and the P\( _{4} \) model is characterized by a strongly
first order transition at temperature 0.376 \cite{abhijit}. In the low temperature
ordered phase in both models the pair correlation function shows an algebraic
decay to a plateau which changes over to an exponential decay in the neighborhood
of the phase transition\cite{enakshi,abhijit}. }{\large \par}

{\large The role of defects in the phase transition of various three dimensional
spin systems is very difficult to study theoretically due to the nonlinearity
introduced in the three dimensional nature of the spins. However simulation
technique may be used to investigate the role of topological defects in these
systems. Lau and Dasgupta\cite{lau} have shown numerically using the conventional
Metropolis algorithm that monopoles (hedgehogs) are necessary for the phase
transition in the three dimensional Heisenberg model. These authors observed
that if one suppresses the formation of these defects in the 3-d Heisenberg
model, the system remains ordered at all temperatures and the transition to
the disordered phase disappears altogether. The present work, which involves
an elaborate MC study, was undertaken to investigate the effect of the suppression
of the topological defects on the phase transitions which the planar P\( _{2} \)
and P\( _{4} \) models exhibit. The work was motivated to a large extent by
the work of Lau and Dasgupta\cite{lau} in the 3-D Heisenberg model. Another
work, along the same line, which must be mentioned in this context, is that
of Lammert et. al.\cite{lam} who, in a MC study, have shown that the nature
of the nematic-isotropic transition in a 3D nematic changes when one suppresses
the formation of the stable line defects, called the disclination lines. Our
work shows that the topological defects play a very crucial role in the phase
transition in the planar P\( _{2} \) and P\( _{4} \) models and although these
models possess the same symmetry they have remarkably different critical behavior.}{\large \par}

{\large In the next section we briefly discuss the nature of the topological
defects in the planar P\( _{2} \) and P\( _{4} \) models and the algorithms
for their identification are presented in the following one. The details of
our MC simulation is then presented followed by the result and discussion.}{\large \par}

\section{{\large Topological defects in the planar models}\large }

{\large In the two dimensional Heisenberg model there exists no stable topological
defects. So the first or the fundamental homotopy group is just the set containing
identity. However in the present planar models (P\( _{2} \) and P\( _{4} \)
) due to the local Z\( _{2} \) symmetry in addition to the O(3) symmetry there
arises stable topological point defects known as 1/2-disclination points, where
the director rotates through an angle of 180\( ^{o} \) around the defect core.
The order parameter space is just the unit sphere S\( ^{2} \) with antipodal
points identified. Any mapping of the other half integral point defects on the
order parameter space is homotopically equivalent to the mapping of the 1/2-defect.
Point defects of integral strength are not stable in these models because of
the so called 'escape to the third dimension'. Any attempt to escape from a
configuration containing a 1/2 point defect leads to a more singular semi infinite
line defect extending from the defect core. So the fundamental or the first
homotopy group of the concerned models is just the two element group Z\( _{2} \)\cite{Mermin1},}{\large \par}

{\large \[
\pi _{1}=\{0,1\}\]
 }{\large \par}

{\large It is known that topological instability does not necessarily imply
physical instability\cite{Mermin1}. If the path connecting the singular to
nonsingular configurations of the free energy involves a configuration of higher
free energy than either, then one may say that the topologically unstable singularity
may possess a considerable degree of physical metastability. This seems to happen
in the P\( _{4} \)-model and may be briefly explained as follows. Consider
a configuration where each of the four spins at the lattice sites which form
a unit square are in a plane and oriented at right angle to their neighbors.
The \( -P_{4}(cos\theta ) \) potential, besides having the global minimum at
\( \theta  \) =0 (or \( \theta  \)=\( \pi  \)), also has a local minimum
at \( \theta =\pi /2 \). If the orientation of the spins are now gradually
changed in order to make \( \theta \rightarrow 0 \) (so as to reach the ground
state ) a potential barrier will have to be overcome and the process becomes
energetically costly. Thus there may exists metastable integral point defects
in this model\cite{enakshi1}. }{\large \par}

{\large The algorithm for the detection of these defects is however nontrivial
as it is not really possible to enclose a 1-defect by four spins alone. The
possible method of detection of these defects is discussed in the next section.
We add that we were unable to detect any such defect because of the low probability
of their formation. }{\large \par}

\section{{\large The Defect finding algorithms}\large }

{\large In order to detect the 1/2-point defects we have followed the algorithm
originally proposed by Vachaspati\cite{tanmay} and subsequently used by others\cite{zapo}.
In order to trace out the topological defects it will be useful to see when
a closed loop in the physical space will enclose a 1/2 disclination point. Let
us consider a triangular plaquette ABC in the physical space. Due to local inversion
symmetry we have to assign antipodal pair points on the unit sphere (S\( ^{2} \))
for each site of the triangular plaquette. Let points corresponding to A, B
and C be (N,S) , (P,P\( ^{1} \)) and (R,R\( ^{1} \)) {[} Fig1{]}. Let us start
from the north pole N. The point P or P\( ^{1} \) is selected depending on
which is closer to N. Let P be the selected point. So the arc NP of the great
circle on the order parameter space is traversed when we go from site A to B
on the physical space. Select from R or R\( ^{1} \) whichever is closer to
P. If the selected point is closer to S then the mapping of ABC is a non contractible
loop and the plaquette will enclose a 1/2-disclination point defect. However
if the selected point for C is closer to N, then the mapping is contractible
to a point on the sphere and no defect will be enclosed by the triangular plaquette.
It may be noted that if S\( _{i} \), S\( _{j} \) and S\( _{k} \) are the
spin variables associated with the points A, B and C then the triangular plaquette
will enclose a 1/2-defect if \begin{equation}
\label{vac}
Sgn[(S_{i},S_{j})(S_{j},S_{k})(S_{k},S_{i})]=-1
\end{equation}
}{\large \par}

{\large Priezev and Pelcovits \cite{pelco} in their work on 3-dimensional nematics
have defined defect counting operators base on this principle. Beside these
two mathematically equivalent methods, an algorithm for detecting 1/2-defects
in RP\( ^{2} \) models can be developed on the method first proposed by Berg
and Luscher\cite{berg}. The method works as follows. The projection matrix
P associated with each unit spin vector \( S \) in RP\( ^{2} \) model obeys
relation P\( ^{2} \) =P and Tr P=1 and its elements may be defined as P\( _{\alpha \beta } \)=\( S_{\alpha }S_{\beta } \)
where \( \alpha  \) ,\( \beta  \) =1,2,3. The charge at a lattice site x\( ^{*} \)
enclosed by an elementary triangular plaquette which has the projection matrices
P\( _{1} \) , P\( _{2} \) and P\( _{3} \) associated with its corners is
given by}{\large \par}

{\large \begin{equation}
\label{berg}
q_{x^{*}}=\frac{1}{2\pi }cos^{-1}\frac{Tr\{P_{3}P_{2}P_{1}\}}{\{TrP_{1}P_{2}TrP_{2}P_{3}TrP_{3}P_{1}\}^{1/2}}
\end{equation}
}{\large \par}

{\large We have used and checked that all the three above mentioned algorithms
for the detection of the 1/2-defects in triangular plaquettes are exactly equivalent
in all cases in both models.}{\large \par}

{\large The detection of the metastable 1-defects in the P\( _{4} \) model
is a non-trivial job and the probability of their formation is very low as this
requires a very 'special arrangement of the order parameter over many uncorrelated
domains' \cite{HIND}. In 3-d nematics, in principle where both line and point
defects (hedgehogs) may form, no point defects are observed in experiments on
quenched nematics \cite{chuang} until the defect network has coarsened appreciably.
It has been observed that the monopoles were only formed by string interactions
and none were generated during the quench. Using the topology, more specifically,
the homology of the order parameter space\cite{HIND}, Hindmarsh has explained
why the expected density of point defects in extremely low. The observation,
briefly speaking, is that in order to cover RP\( ^{2} \) space twice (which
is necessary for a topological 1-charge) a roughly spherical arrangement of
a minimum of twelve uncorrelated adjacent domains is necessary and this has
a probability \textasciitilde{}10\( ^{-8} \). In a two dimensional nematic
the same consideration are believed to apply for the the +1 point defects\cite{hindmarsh}.
We have used the algorithm using 12-spin configuration proposed by Zapotacky
et. al. \cite{zapo} to detect the 1-charge but could find none. }
\begin{figure}
{\par\centering \resizebox*{0.8\textwidth}{!}{\includegraphics{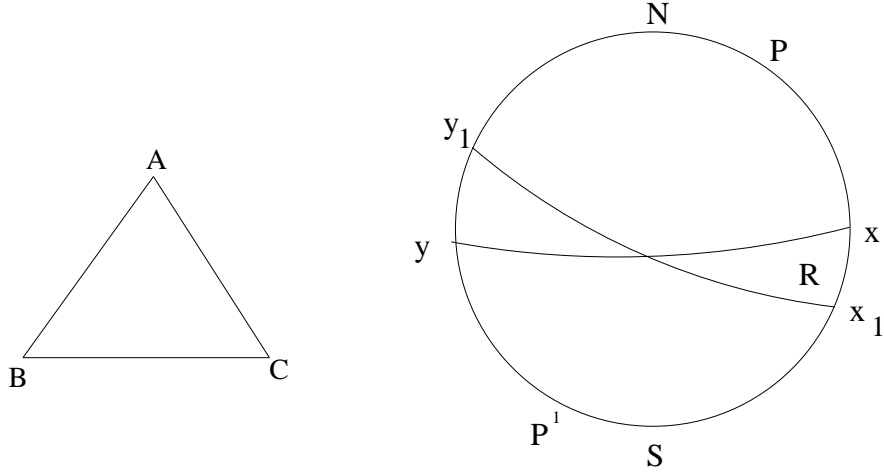}} \par}

\caption{The Triangular plaquette ABC in physical space (left) and the order parameter
space (right). On S\protect\( ^{2}\protect \) (N,S) is assigned for A, (P,P\protect\( ^{1}\protect \))
for B and (R,R\protect\( ^{1}\protect \)) for C. The great circle xy is perpendicular
to NS and x\protect\( _{1}\protect \)y\protect\( _{1}\protect \) is perpendicular
to PP\protect\( ^{1}\protect \). If the point R is outside the region enclosed
by two great circles then the corresponding loop is contractible, otherwise
ABC will enclose a 1/2-disclination point. {[}see ref.\cite{tanmay}{]}}
\end{figure}
{\large \par}

\section{{\large The Simulation Details}\large }

{\large In the present paper we have used the conventional Metropolis spin update
algorithm\cite{metro,Barkema} with periodic boundary condition in order to
study the role of the topological defects in the phase transitions exhibited
by the planar P\( _{2} \) and P\( _{4} \) model. Lattice sizes ranging from
20x20 to 80x80 were used and a part of the work was performed using the histogram
reweighting technique of Ferrenberg and Swendsen \cite{swe}.}{\large \par}

{\large In order to carry out the procedure of the suppression of the defects
in the planar lattice models we have included a chemical potential term associated
with the topological charges \cite{lau}. The Hamiltonian used in the simulation
is given by,\begin{equation}
\label{ham}
H=-\sum _{ij}P_{L}(cos\theta _{ij})+\lambda \sum _{ijkl}Q_{ijkl}
\end{equation}
 where L is either 2 or 4. \( \theta _{ij} \) is the angle between the nearest
neighbor spins i, j and the \( Q_{ijkl} \) is the sum of charges of two triangular
portions of a square plaquette. A positive \( \lambda  \) makes the formation
of the charges expensive in terms of energy and for almost total suppression
of the charge a large value of \( \lambda  \) (about 10 to 60, but independent
of temperature) was normally chosen. In order to obtain the unrestricted simulation
we set \( \lambda =0 \). The charge enclosed by ijk for instance is given by.}{\large \par}

{\large \begin{equation}
\label{charge}
Q_{ijk}=\frac{1}{4}[1-sgn\{(S_{i},S_{j})(S_{j},S_{k})(S_{k},S_{i})\}]
\end{equation}
 }{\large \par}

{\par\raggedright {\large Clearly the sum \( Q_{ijkl} \) can be 0, 1/2 or 1.
If \( Q_{ijkl} \) is 0 then the square plaquette encloses no charge. If it
is 1/2 then a 1/2-disclination point is enclosed. But if it is 1, then it should
not be confused with an integral point defect. In fact this corresponds to two
closest possible 1/2-charges situated within a square plaquette of linear dimension
equal to lattice spacing. }\large \par}

{\large In both the planar models we have investigated, the smallest part of
the system in real space that can enclose a 1/2-point defect is a triangle.
Each elementary square plaquette can be diagonally cut into two triangles and
if these two adjacent triangles each enclose a 1/2-defect then this leads to
a 1/2-pair charge being enclosed by the elementary square. If only one of those
triangles encloses a 1/2-charge then the square in turn encloses a single 1/2-charge.
We denote number of square plaquettes enclosing a pair of 1/2-charges by n\( _{1} \)
(this should however not be confused with a topological defect of charge 1).
Similarly the number of elementary squares enclosing a single 1/2-charge is
denoted by n\( _{1/2} \). In a recent work, Mondal and Roy\cite{enakshi1}
have observed that the ratio n\( _{1/2} \)/n\( _{1} \) behaves like a response
function in both the models although their behavior is different in the two
systems. When plotted against temperature, the ratio n\( _{1/2} \)/n\( _{1} \)
exhibits a maximum at the transition in the P\( _{2} \) model while in the
P\( _{4} \) model it shows a sharp fall at the transition. Finite size effects
are also prominent in the transition temperatures thus obtained from the n\( _{1/2} \)/n\( _{1} \)
vs T plots. }{\large \par}

{\large In our MC simulation, while investigating the effect of suppression
for the charges in the two planar models, we have treated the single 1/2-charge
and the 1/2-pairs (within an elementary square plaquette) on different footings.
For the simulation where no charge suppression has been attempted we have set
\( \lambda =0 \). For the suppression of the charges represented by n\( _{1/2} \)
, \( \lambda \neq 0 \) only for Q\( _{ijkl} \) =1/2 while for total (both
single-1/2 and 1/2-pair suppression), \( \lambda \neq 0 \) for Q\( _{ijkl}\neq 0 \).
In the P\( _{2} \) model, complete suppression of single 1/2-defects was found
to lead to complete suppression of the 1/2-pair defects and this leads to complete
elimination of the phase transition in this model. In the P\( _{4} \) model,
however, the suppression of the single 1/2-defects leaves a large number of
1/2-pair defects within the elementary squares and the evidence of a new phase
transition at a higher temperature is obtained. When these defects too were
suppressed, the phase transition totally disappears. }{\large \par}

{\large In order to estimate the critical exponents and the thermodynamic limit
of the critical temperature of the new transition which we obtained in the P\( _{4} \)
model and which seems to be of second order, we have applied the finite size
scaling method. Finite size scaling method is a technique of estimating the
critical exponents and the thermodynamic limit of the transition temperature
by observing how the measured quantities vary with the system size. In finite
size scaling method (the data collapse method in particular) we extract the
part of the thermodynamic function which does not contain system size explicitly\cite{Barkema}.
This part is called the scaling function. If proper values of the critical exponents
and the thermodynamic limit for the transition temperature are chosen then the
scaling function for different system sizes get collapsed. In this paper we
have used the data collapse technique for estimating the critical exponents
associated with the specific heat and the order parameter. The critical exponents
associated with the correlation length, specific heat and the order parameter
are denoted by \( \nu  \), \( \alpha  \) and \( \beta  \) respectively. }{\large \par}

{\large For specific heat, the scaling relation stands as}{\large \par}

{\large \begin{equation}
\label{cv}
C_{V}=L^{\alpha /\nu }\widetilde{C}(L^{\frac{1}{\nu }}t)
\end{equation}
}{\large \par}

{\large where t is the reduced temperature and \( \widetilde{C} \) is the specific
heat scaling function. }{\large \par}

{\large Similarly for the order parameter ( <q\( ^{2} \)>)(see eq. \ref{ord}),
the scaling relation is given by}{\large \par}

{\large \begin{equation}
\label{lamm}
<q^{2}>=L^{-4\beta /\nu }\widetilde{Q}(L^{\frac{1}{\nu }}t)
\end{equation}
}{\large \par}

{\large where \( \widetilde{Q} \) is known as the order parameter scaling function\cite{lam}.}{\large \par}

\section{{\large RESULTS AND DISCUSSION}\large }

{\large We have evaluated various thermodynamic properties like internal energy
per particle (<E>), specific heat, order parameter etc. The specific heat was
evaluated by taking the temperature derivative of the <E> as well as from fluctuation
of the energy. Due to the local inversion symmetry, the order parameter is a
second rank tensor. }{\large \par}

{\large As a measure of the order prevailing in the system we used the quantity
<q\( ^{2} \)> given in ref.\cite{lam}}{\large \par}

{\large \begin{equation}
\label{ord}
<q^{2}>=\frac{N}{N-1}<\frac{3}{2}TrQ^{2}-\frac{1}{N}>
\end{equation}
}{\large \par}

{\par\raggedright {\large where \( Q_{ab}=\frac{1}{N}\sum Q(i)_{ab} \) is the
nematic tensor order parameter, where \( Q(i)_{ab}=(n_{a}n_{b}-\frac{1}{3}\delta _{ab}) \)
(\( \widehat{n} \) is the molecular axis of the} \noun{\large \( i^{th} \)}
{\large molecule) and N is the total number of sites (i) in the lattice. This
definition ensures that <q\( ^{2} \)> is zero in a fully disordered system
and 1 for a fully ordered system. }\large \par}

{\large In case of P\( _{2} \) model we have simulated for linear dimension
L=40 and 60. In Fig2 we have depicted the specific heat versus temperature plot
for the P\( _{2} \) lattice model of size 40x40. The unrestricted simulation
shows a peak which disappears when the single 1/2-charges on the square plaquettes
were suppressed. We have observed that the after suppression of the single 1/2-charges
on the square plaquettes there remains no 1/2-pair defect in any square plaquette.
The temperature dependence of the order parameter used in this model for the
two cases are shown in Fig3. While the temperature derivative of <q\( ^{2} \)>
has a peak at the transition temperature in the normal case (where no defect
is suppressed), which presumably is a signal of a phase transition in a finite
system \cite{q2}, in the defect free case it seems to lose the characteristic
shape and shows a smooth and rather slow decrease with temperature and vanishes
at around a temperature T=6. We would be inclined to conclude from the results
on C\( _{V} \) and <q\( ^{2} \)> that the defect free phase exhibits no phase
transition at all. }{\large \par}

{\large Turning to the P\( _{4} \) model, which has a characteristic strongly
first order phase transition\cite{abhijit}, we first point out that the suppression
of the single 1/2 defects on the square plaquettes here does not result in suppression
of the 1/2-pair defects on the square plaquettes. This observations is different
from what happens in the P\( _{2} \) system where the suppression of the single
1/2 defects on the square plaquettes leads to suppression of the 1/2-pair defects.
On suppressing the single 1/2-defects a new phase transition is observed. We
however point out that it is impossible to make the system completely free of
topological defects, even when arbitrarily large values of \( \lambda  \) are
used. However the residual charges left were of very insignificant amount. For
instance, at T=0.55, the traces of the single 1/2 and 1/2-pair charges which
could not be suppressed were about 0.02\% of these charges present in the system
at the same temperature after single 1/2-charge suppression. Fig4 shows the
temperature dependence of the specific heat for the normal lattice (L=40) and
after suppression of the single 1/2-defects ( for L=40 and 60). The peak in
C\( _{V} \) for the latter case, while greatly reduced in size (from 80 to
about 17), shifts to a higher temperature which for both lattices is close to
0.494. Presumably the phase transition which we observed after suppressing the
single 1/2 defects is due to the presence of the 1/2-pair defects. When we suppressed
the 1/2-pair defects too in all triangular plaquettes of the lattice no evidence
of any peak in C\( _{V} \) was observed, although a hump like feature was seen
with C\( _{V} \) \textasciitilde{} 3.5 over a temperature range extending from
0.53 to 0.62. In case of P\( _{4} \) model we have also investigated the response
of the defect density to the phase transition. When no kind of charge suppression
was applied a large number of single 1/2 defects as well as 1/2-pair defects
were found to be present. In fig5 the temperature dependence of the density
of single 1/2-defects are shown for unrestricted simulation. In fig6 the temperature
dependence of the density 1/2-pair defects is shown for both before and after
suppression of the single 1/2 defects. Clearly the temperature derivative of
defect density behaves like other response functions for all the cases.}{\large \par}

\begin{figure}
{\par\centering \resizebox*{0.85\textwidth}{!}{\includegraphics{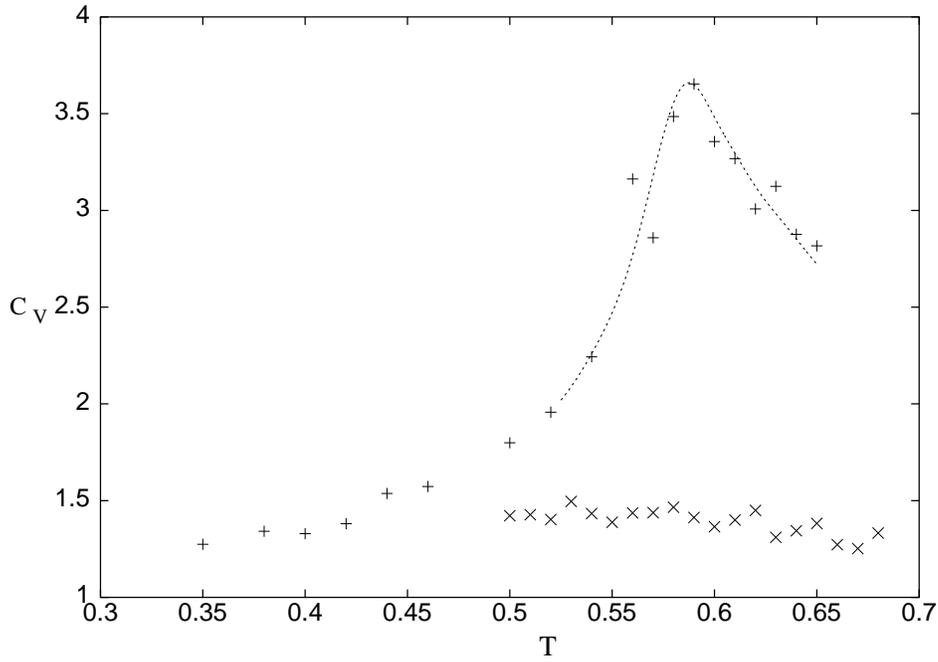}} \par}

\caption{The specific heat, vs. temperature plot in P\protect\( _{2}\protect \) model.
+ corresponds to C\protect\( _{V}\protect \) obtained from energy fluctuation
in the L=40 lattice and the continuous curve is taken from ref\cite{enakshi}
where it has been obtained using multiple histogram reweighting and peak is
at T=0.587. Both corresponds to normal MC simulation with no charge suppression.
The x represents C\protect\( _{V}\protect \) for L=40 lattice after suppressing
the single 1/2 defects on the square plaquettes. }
\end{figure}

\begin{figure}
{\par\centering \resizebox*{0.8\textwidth}{!}{\includegraphics{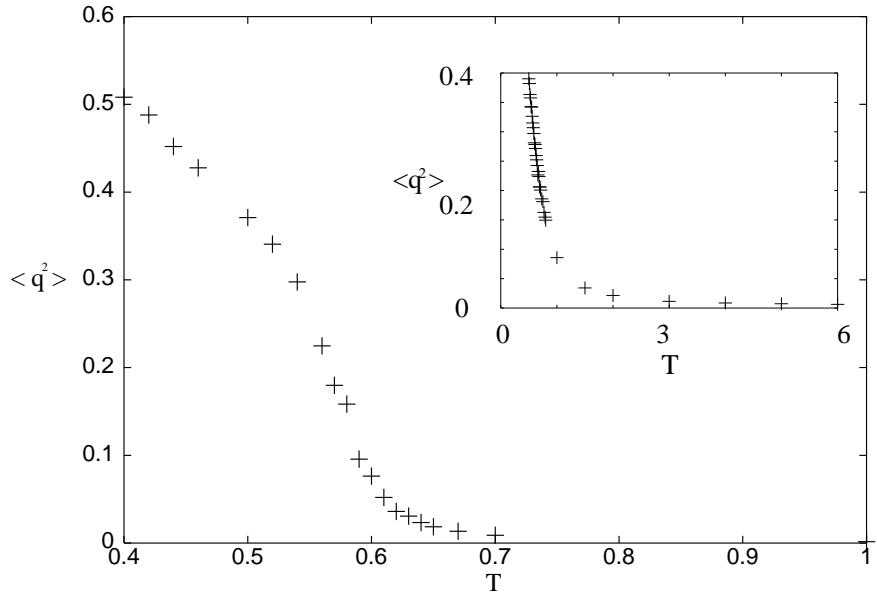}} \par}

\caption{The order parameter <q\protect\( ^{2}\protect \)> obtained for the L=40 lattice
in the P\protect\( _{2}\protect \) model with no charge suppression. The temperature
derivative has a peak at T=0.567. The inset shows the <q\protect\( ^{2}\protect \)>
vs T plot after suppressing the single 1/2-defects on the square plaquettes.}
\end{figure}

\begin{figure}
{\par\centering \resizebox*{0.8\textwidth}{!}{\includegraphics{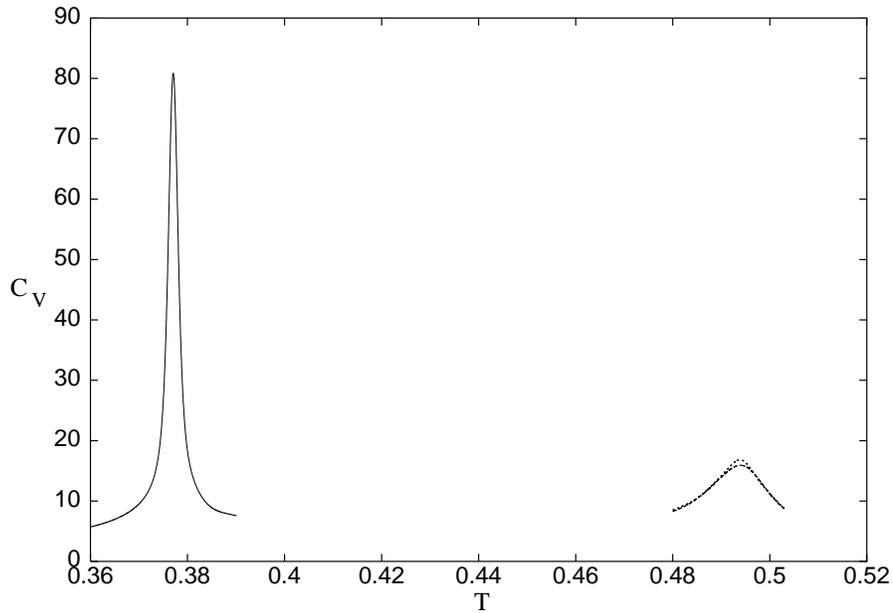}} \par}

\caption{C\protect\( _{V}\protect \) plotted against T for unrestricted simulation
of L=40, P\protect\( _{4}\protect \) lattice (left) and the same for L=40 and
L=60 lattice obtained after suppressing the single 1/2-defects on the square
plaquettes. The slightly bigger peak is for the L=60 lattice. The peak occurs
at T=0.494.}
\end{figure}

\begin{figure}
{\par\centering \resizebox*{0.8\textwidth}{!}{\includegraphics{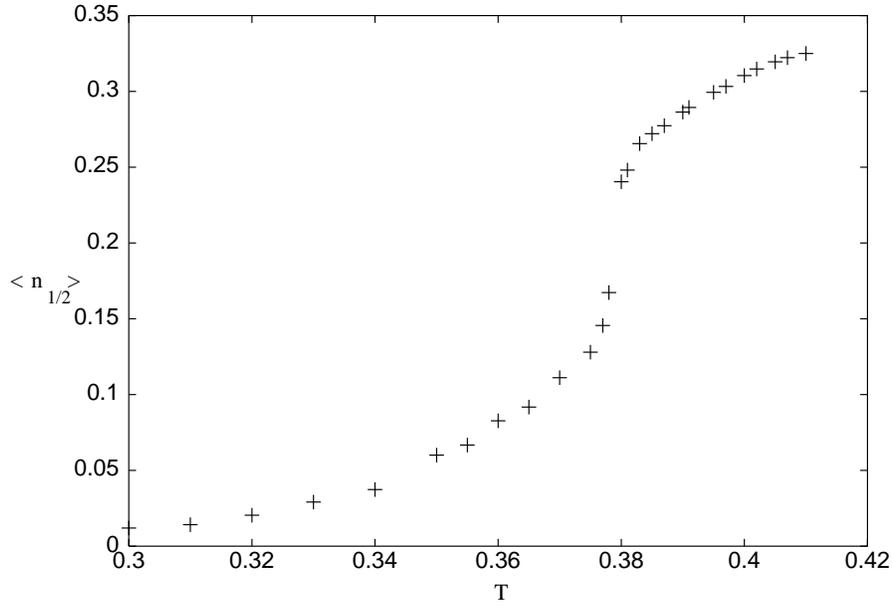}} \par}

\caption{The density of the single 1/2-defects enclosed by the square plaquettes in
case of L=40 P\protect\( _{4}\protect \) model with no charge suppression.}
\end{figure}

\begin{figure}
{\par\centering \resizebox*{0.85\textwidth}{!}{\includegraphics{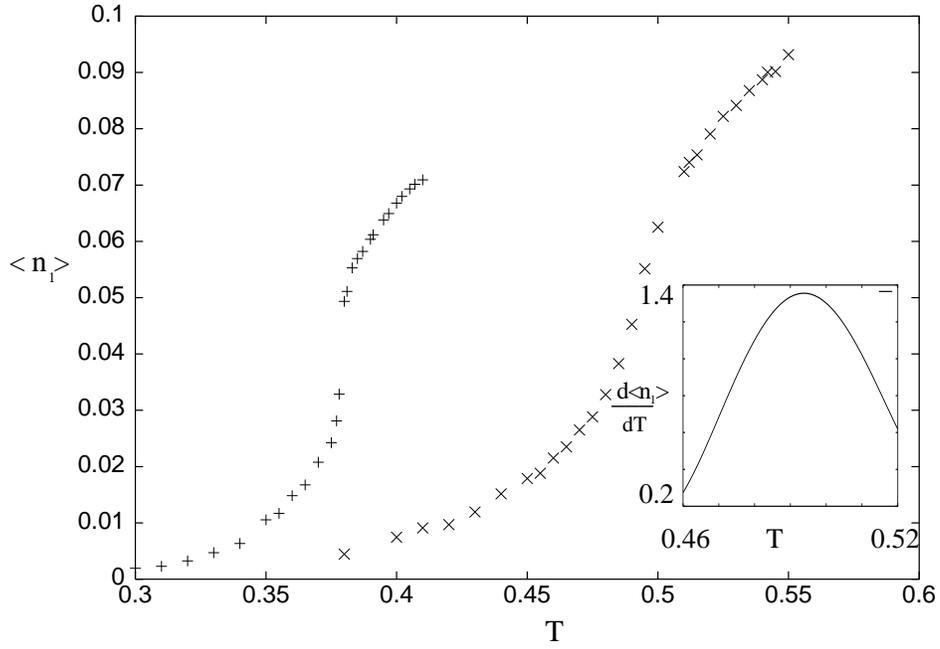}} \par}

\caption{The density <n\protect\( _{1}\protect \) > of 1/2-pair charges for unrestricted
case of L=40 P\protect\( _{4}\protect \) model (left) and the same after suppressing
the single 1/2-defects . The temperature derivative of the former has a peak
at T=0.379 and the latter(inset) peaks at T=0.494.}
\end{figure}

\begin{figure}
{\par\centering \resizebox*{0.85\textwidth}{!}{\includegraphics{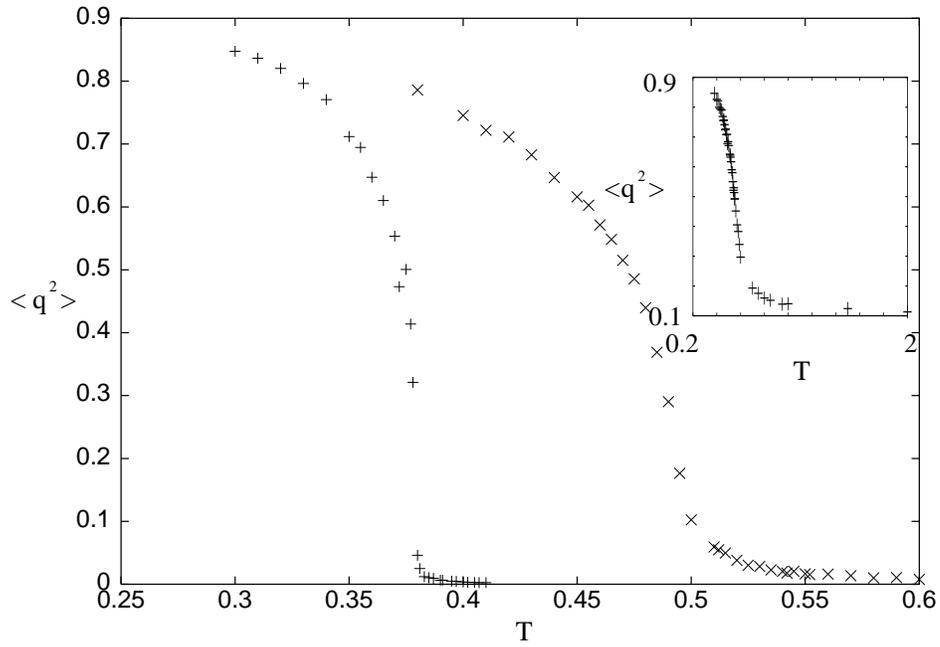}} \par}

\caption{The temperature dependence of <q\protect\( ^{2}\protect \) > for the L=40
P\protect\( _{4}\protect \) model. On the left is the unrestricted MC result
and on the right is the case when the single 1/2-charge on square plaquette
were suppressed. The inset shows <q\protect\( ^{2}\protect \)> when all charges
are suppressed. The temperature derivative of the first two curves have peaks
at 0.379 and 0.490 respectively.}
\end{figure}

\begin{figure}
{\par\centering \resizebox*{0.85\textwidth}{!}{\includegraphics{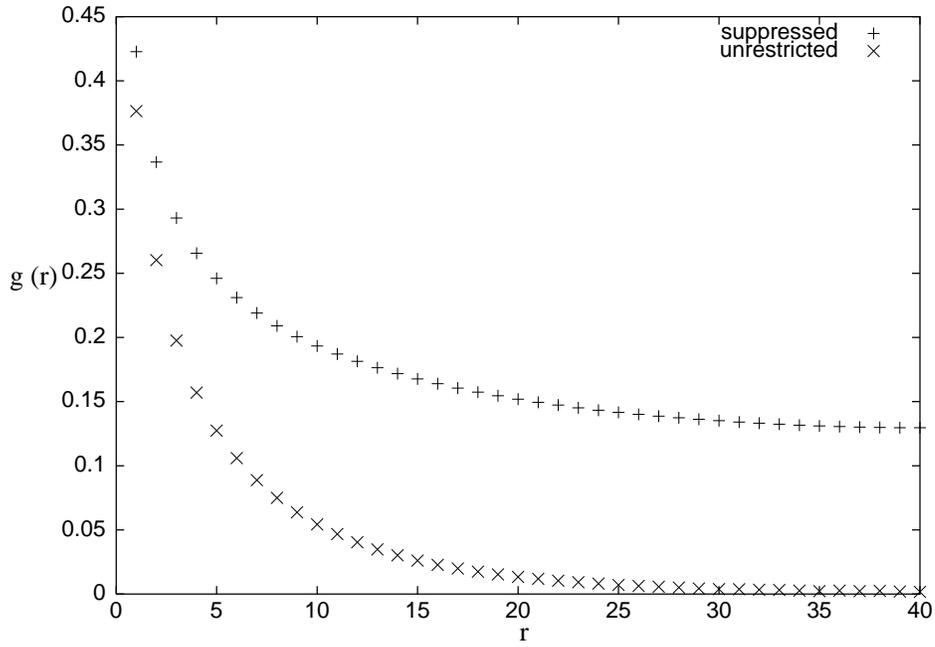}} \par}

\caption{The correlation function g(r) plotted against r for L=40 P\protect\( _{2}\protect \)
lattice at T=0.6 for both unrestricted case and the suppressed case.}
\end{figure}

\begin{figure}
{\par\centering \resizebox*{0.85\textwidth}{!}{\includegraphics{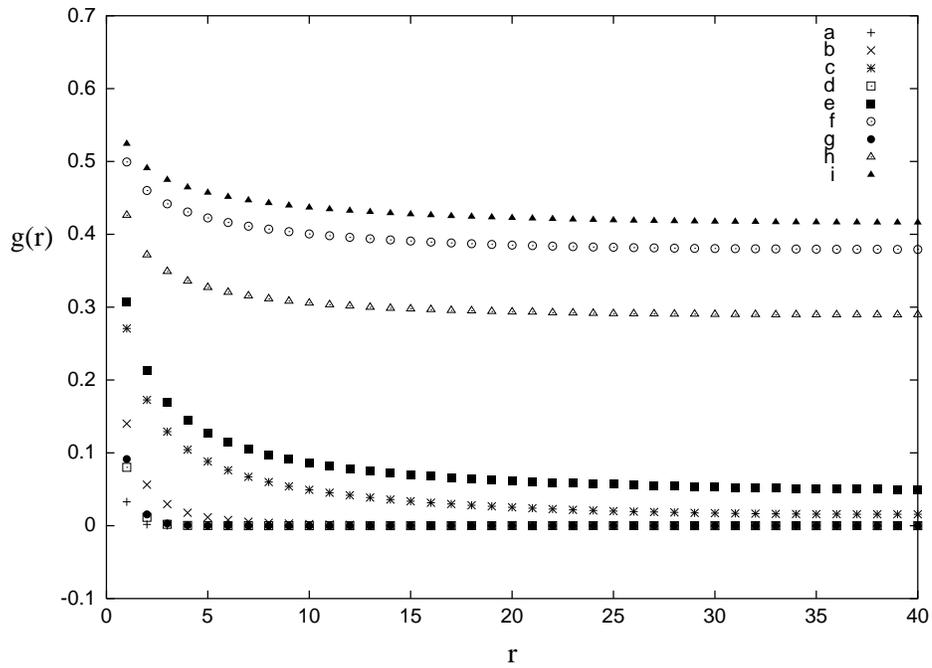}} \par}

\caption{The correlation function g(r) plotted against r for the L=80, P\protect\( _{4}\protect \)
lattice. For the curves (a), (b), (c), T=0.68, for the curves (d), (e), (f),
T=0.5 and for (g), (h), (i), T=0.48. The three curves in each set correspond
to the normal MC, 1/2-charge suppressed and total charge suppressed cases. The
parameters used to fit these curves are listed in Table1.}
\end{figure}

\begin{figure}
{\par\centering \resizebox*{0.85\textwidth}{!}{\includegraphics{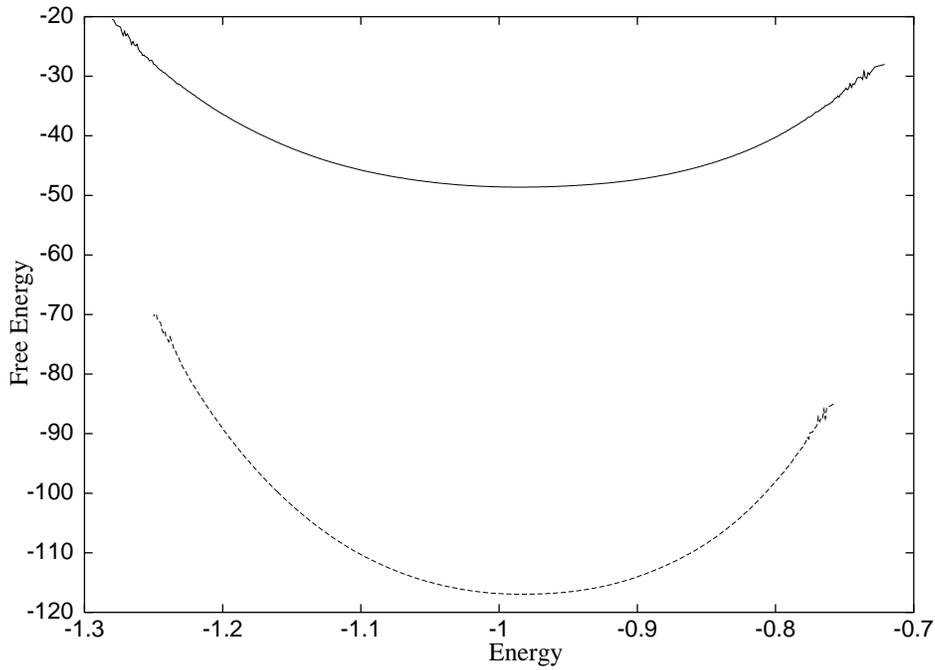}} \par}

\caption{Free energy vs energy for P\protect\( _{4}\protect \) model for L=40(upper)
and L=60(lower) sizes after suppressing the single 1/2-defects on the square
plaquettes.}
\end{figure}

\begin{figure}
{\par\centering \resizebox*{0.85\textwidth}{!}{\includegraphics{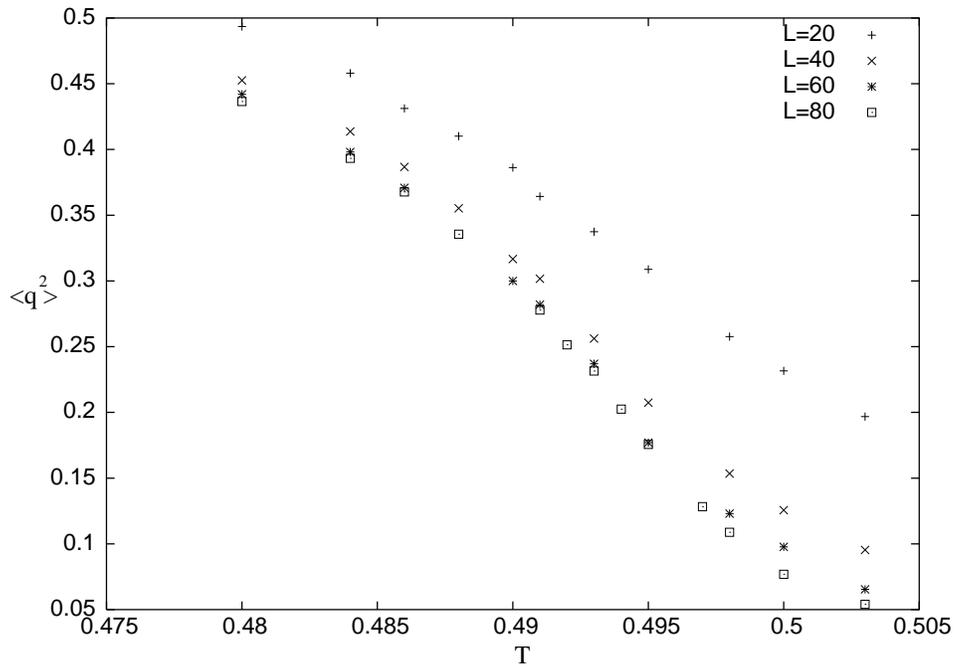}} \par}

\caption{The order parameter <q\protect\( ^{2}\protect \)> plotted against temperature
T for three lattice sizes indicated for the P\protect\( _{4}\protect \) model
after suppressing single 1/2-defects.}
\end{figure}

\begin{figure}
{\par\centering \resizebox*{0.8\textwidth}{!}{\includegraphics{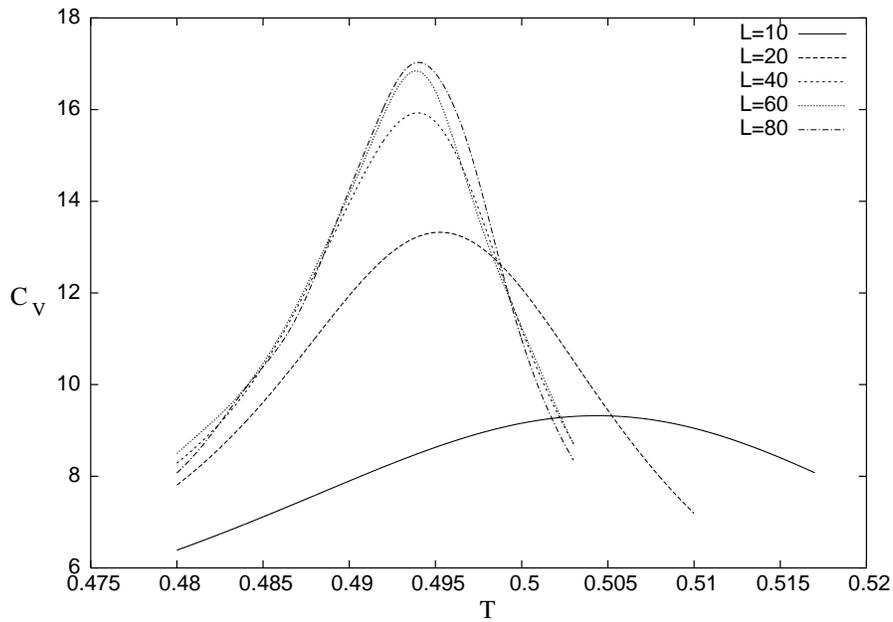}} \par}

\caption{The specific heat C\protect\( _{V}\protect \) vs T for the lattice sizes indicated
for the P\protect\( _{4}\protect \) model after the single 1/2-charge suppression.
The values of C\protect\( _{V}\protect \) here obtained from d<E>/dT and the
histogram reweighting technique \cite{swe} was used.}
\end{figure}

\begin{figure}
{\par\centering \resizebox*{0.8\textwidth}{!}{\includegraphics{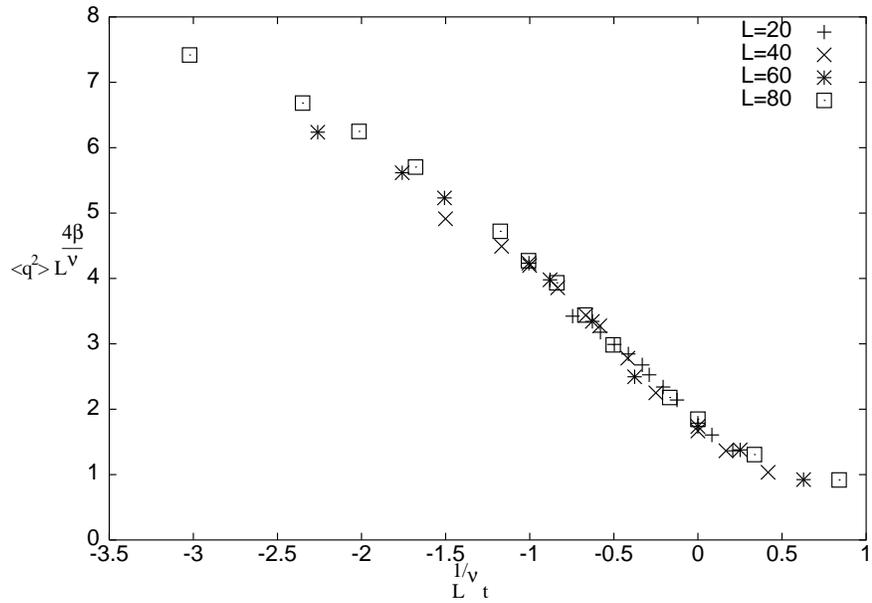}} \par}

\caption{Collapse of the order parameter for L=20,40,60 and 80 sizes. The best collapse
obtained at Tc(\protect\( \infty \protect \)) = 0.498 (thermodynamic limit),
\protect\( \nu =\protect \)0.99 (exponent for correlation length), \protect\( \beta =\protect \)0.16
(exponent for order parameter). t is the reduced temperature, (T-T\protect\( _{c})\protect \)
/T\protect\( _{c}\protect \)}
\end{figure}

\begin{figure}
{\par\centering \resizebox*{0.85\textwidth}{!}{\includegraphics{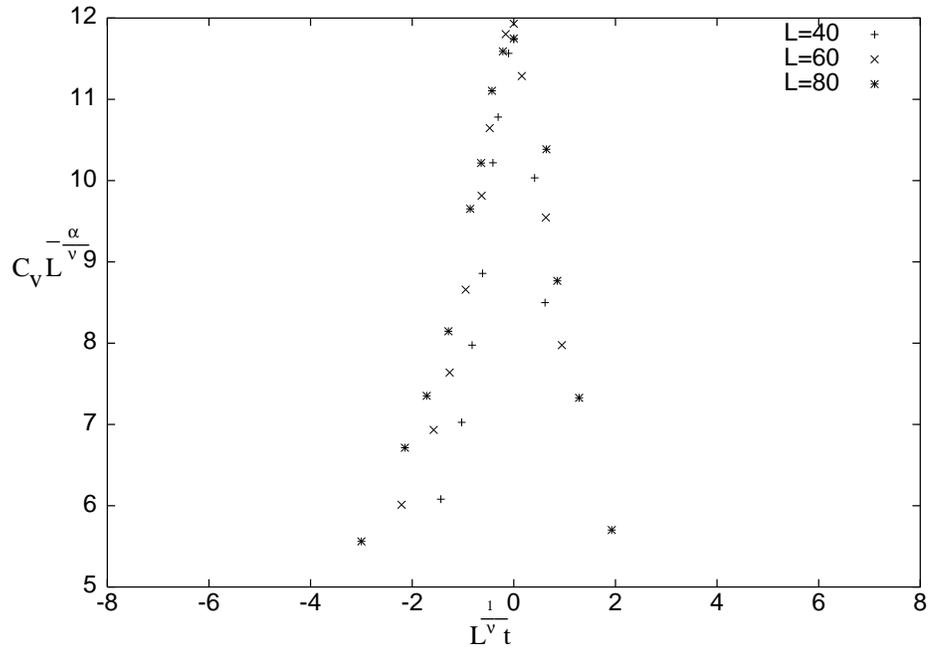}} \par}

\caption{The collapse of the specific heat data for system sizes L=40, 60 and 80. The
best collapse obtained at thermodynamic limit of the transition temperature
(Tc(\protect\( \infty \protect \))=0.4938), \protect\( \nu \protect \) = 0.94
and \protect\( \alpha \protect \)(exponent for the specific heat) = 0.077. }
\end{figure}

\begin{figure}
{\par\centering \resizebox*{0.85\textwidth}{!}{\includegraphics{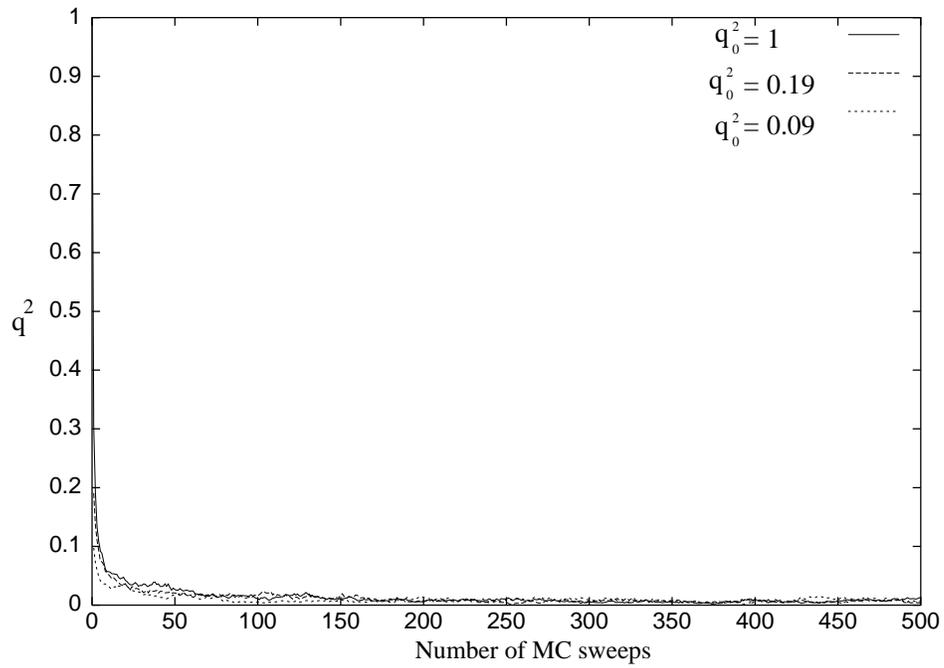}} \par}

\caption{Evolution of order parameter for 80x80 P\protect\( _{2}\protect \) model after
suppressing the 1/2-defects enclosed by the square plaquettes using \protect\( \lambda \protect \)
=60. The final values of order parameter for three initial states with q\protect\( _{0}\protect \)\protect\( ^{2}\protect \)
=1.0, q\protect\( _{0}\protect \)\protect\( ^{2}\protect \) =0.19 and q\protect\( _{0}\protect \)\protect\( ^{2}\protect \)=0.09
are almost same.}
\end{figure}

\begin{figure}
{\par\centering \resizebox*{0.9\textwidth}{!}{\includegraphics{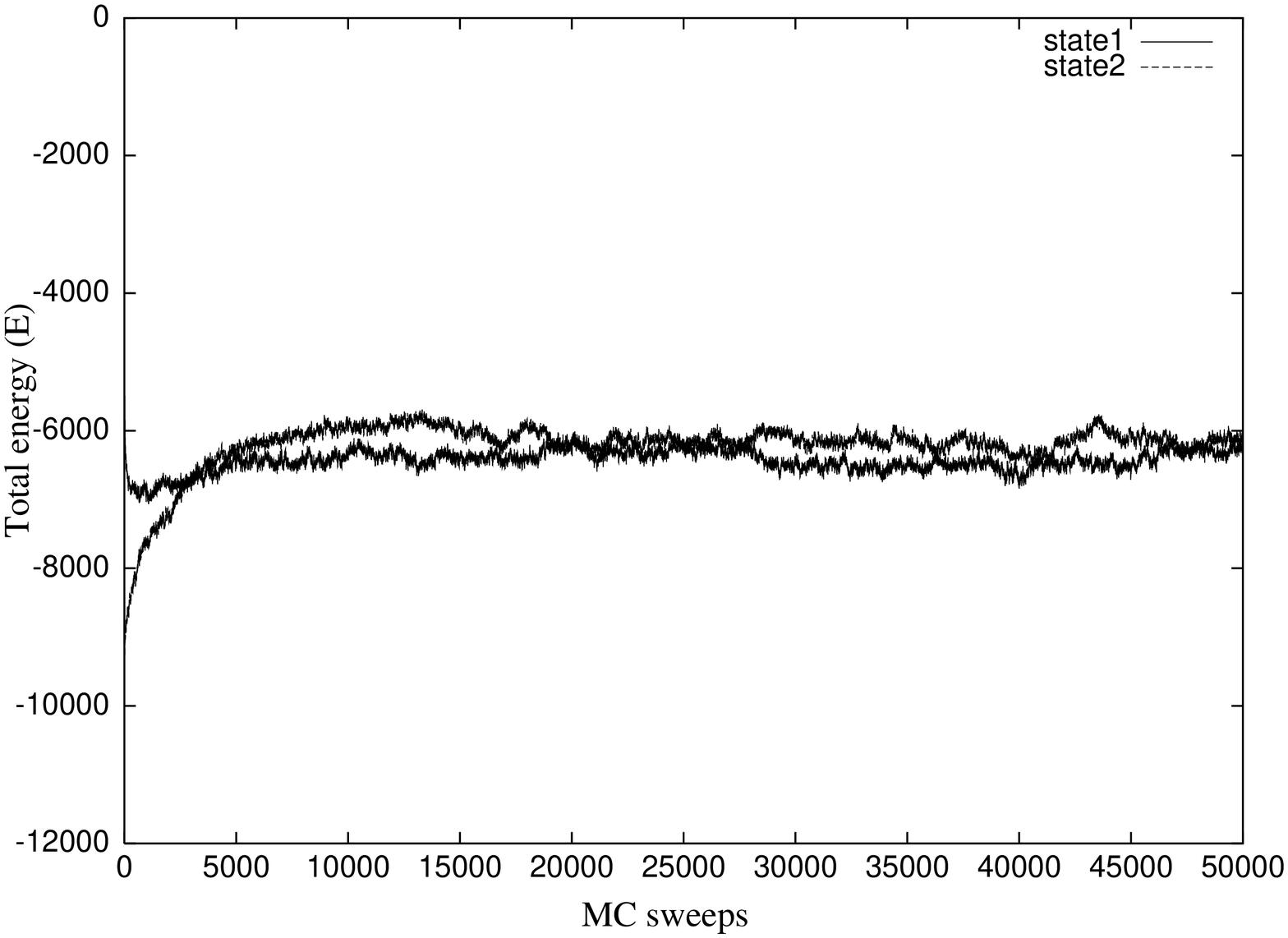}} \par}

\caption{Evolution of total energy of the 80x80 P\protect\( _{4}\protect \) model after
suppressing the 1/2-defects enclosed by the square plaquettes using \protect\( \lambda \protect \)=60.
The final values of Energy for two different initial states are almost same. }
\end{figure}

{\large In figure 7, we have depicted the temperature dependence of the order
parameter <q\( ^{2} \)> for all three cases in the P\( _{4} \) model for L=40.
The peaks of temperature derivative of <q\( ^{2} \)> occur respectively at
0.379 (normal MC) and 0.490 (after suppressing single 1/2-defects) and their
heights are 119.5 and 18.5 respectively. On suppressing all the 1/2 charges
on all triangular plaquettes of the lattice (total charge suppression), d<q\( ^{2} \)>/dT
no longer exhibits a peak and <q\( ^{2} \)> seems to have an asymptotic value
of 0.1 at T=2, up to which the investigation has been made. We add that d<q\( ^{2} \)>/dT
after total charge suppression has a feature similar to that seen in C\( _{V} \)
in that a broad peak of height \textasciitilde{}4 was seen over a temperature
range from 0.52 to 0.6. The existence of this broad hump of insignificant magnitude
over an extended temperature range can not be a sign of phase transition. These,
in someway, may be connected to the existence of the small number of residual
charges left in the system after the attempt to suppress them completely failed.}{\large \par}

{\large We now turn to the pair correlation function g(r)\( =<cos^{2}\theta _{ij} \)\( - \)\( 1/3> \)
where r is the separation between the two spins i and j which make an angle
\( \theta  \)\( _{ij} \) with each other. This function for the P\( _{2} \)
model at T=0.6, for instance, which is a temperature much higher than the normal
critical temperature in this model, decays exponentially to zero, as it should,
in the complete absence of long range order and a best fit with g(r)\( =\alpha exp(-\lambda r) \)
yields \( \alpha  \)=0.439 and \( \lambda  \)=0.248. For the single 1/2-charge
(and consequently of 1/2-pair charges) suppressed case, it decays algebraically
and a best fit like g(r)\( =ar^{-p}+b, \) yields the parameters a=0.404, b=0.022
and p=0.372 {[}Fig 8{]}. In figure 9, the g(r) vs r plots for the P\( _{4} \)
model are shown for T=0.48 (which is greater than the normal transition temperature
but less than the transition temperature obtained after 1/2-charge suppression),
T=0.5 (which is slightly higher than the observed transition temperature after
1/2-charge suppression) and T=0.68. At all of these temperatures, as one would
expect, the correlation function in the normal case decays exponentially to
zero. The two other cases give best fits for an algebraic decay to a plateau,
and the parameters are listed in the Table1. We find that in the P\( _{4} \)
model the results of single 1/2-charge suppression is the same as that in the
P\( _{2} \) model while with the suppression of the 1/2-pair charges too, the
asymptotic value of the order prevailing in the system increases further. The
phase transition, which the P\( _{4} \) system has after suppressing the single
1/2-charges is second order. We have evidence that it is not first order as
the dual peak structure of the probability distribution as a function of energy,
which is so distinct in the P\( _{4} \) model\cite{abhijit}, has been found
to disappear totally after suppressing the single 1/2-charges on the square
plaquettes. In fig10 we have given the the free energy of 40x40 and 60x60 lattice
in P\( _{4} \) model after suppressing the single 1/2 defects on the square
plaquettes. The single well structure of the free energy indicates the second
orderdness of the new phase transition after suppressing the single 1/2-defects. }{\large \par}

{\large We have also used standard finite size scaling method available for
second order phase transition\cite{Barkema} in order to estimate the critical
exponents and the thermodynamic limit of transition temperature of the new phase
transition that was observed after suppressing the single 1/2-defects. In Fig11
and Fig12 we have depicted the order parameter <q\( ^{2} \)> and the specific
heat C\( _{V} \) plots respectively as function of temperature for different
lattice sizes after the single 1/2-charge suppression in the P\( _{4} \) model.
We have used standard data collapse technique to collapse the data of the figures11
and 12 and the resulting diagrams are Fig13 and Fig14 respectively. In fig13
we have shown the collapse of the order parameter scaling function for L=20,
40, 60 and 80. While in fig14 the collapse of specific heat scaling function
for the system sizes 40,60 and 80 are displayed. The data collapse clearly shows
the phase transition that P\( _{4} \) model exhibits after the single 1/2-charge
suppression is second order. The parameters we have obtained are T\( _{c} \)=0.498,
\( \beta = \)0.16 and \( \nu  \)=0.99 for the collapse of order parameter
and \( T_{c} \) =0.4938 , \( \alpha = \)0.077 and \( \nu = \)0.94 from collapse
of the specific heat. We would remark that the high temperature behavior of
the order parameter <q\( ^{2} \)> after total charge suppression does not indicate
the presence of a phase transition. This is true for both models. }{\large \par}

{\large We have used large values of \( \lambda  \) in our simulation in order
to suppress the evolution of defects in both models. It is known that any Monte
Carlo study is faithful only if we can reach any point in the phase space starting
from any other point . So there must be a path connecting the two points in
phase space with non zero probability. This actually indicates that we should
be careful that we are not trapped in any small region of the phase space. We
have investigated the phase space connectivity in both the models by observing
the evolution of order parameter or energy with MC steps. The connectedness
is satisfied if the observed quantities for different initial states converge
to the same final value. In fig15 we have shown that in case of the 80x80 P\( _{2} \)
model, after suppressing the 1/2-defects ( by using \( \lambda  \)=60 ) on
the square plaquettes, the final values of the order parameter is same for three
different initial configurations. Similarly in Fig16 we have shown that the
same thing happens for the total energy of the 80x80 P\( _{4} \) model after
suppressing the 1/2-defects on the square plaquettes. It is therefore clear
that we can use a value of \( \lambda  \) at least up to 60 without violating
the phase space connectivity.}{\large \par}

\section{CONCLUSION}

{\large It is established in this paper that topological defects play a very
important role in the phase transitions exhibited by the two planar lattice
models we discussed. The observed difference in the critical behavior seems
to be due to the difference in the role played by the topological defects. It
is shown in this paper that for the phase transition in both the models topological
defects are necessary. In the P\( _{2} \) model the phase transition is governed
by the 1/2-disclination points enclosed by the square plaquettes only (single
1/2). On suppressing these single 1/2-defects we have shown that the phase transition
was totally eliminated. However the picture is different in case of two dimensional
P\( _{4} \) model. Where on suppressing the single 1/2-defects the nature and
the transition temperature of the phase transition gets changed. We have also
shown that the new phase transition in the P\( _{4} \)-model is due to the
presence of large number of 1/2-pairs (enclosed by two triangular portions )
within the square plaquettes which remains unsuppressed even after suppressing
the single 1/2-defects. This leads to another important conclusion that for
tracing out all the disclination points in two dimension, the minimum closed
loops in physical space to be considered are the smallest triangular cells formed
by three nearest neighbor sites. }{\large \par}

{\large It will be interesting to apply the idea of triangular plaquettes in
three dimensional nematics, where the stable topological defects are disclination
lines as well as monopoles. Let us consider small disclination loop in a three
dimensional nematic {[}fig17{]}. In the figure a cubic unit cell ABCDEFGH is
shown. The disclination loop has crossed the face ABCD twice. However if we
take mapping of ABCD on the order parameter space then we would get a contractable
loop in the order parameter space and conclude that ABCD does not enclosed any
line defect. It is true that ABCD does not enclose a single line, but it encloses
two close lines (as the loop intersects the face ABCD twice), which could be
detected only if we consider the two triangular portions (ABC and ADC) of ABCD.
So considering the elementary plaquettes to be square plaquettes of linear dimension
equal to lattice spacing, we would miss loops as shown in the figure. In real
nematics small loops are known to carry monopole (hedgehog) charges. Actual
monopole structure is very rare in real nematics. As already stated, the low
probability of actual monopole structure is discussed by Hindmarsh \cite{HIND}.
The monopoles that comes in to play in the three dimensional nematics arises
mainly from the small disclination rings. In our work it is evident that in
P\( _{4} \) model it is very important to consider these triangular plaquettes
in order to trace out the topological defects properly. If the same work is
carried out on three dimensional P\( _{4} \) model and disclination lines were
traced out using triangular plaquettes, then one is likely to get a large number
of rings as shown in the figure17. The small disclination rings which behave
like monopoles, is expected to have important role in the phase transition of
the three dimensional P\( _{4} \) model. }{\large \par}

\begin{figure}
{\par\centering \resizebox*{0.4\textwidth}{!}{\includegraphics{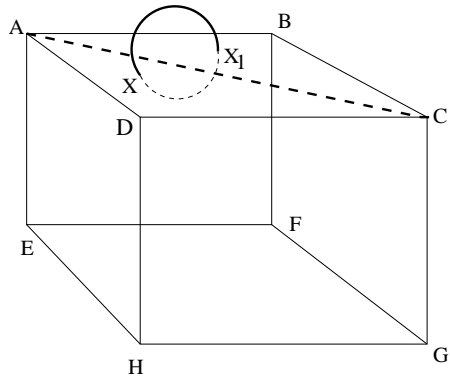}} \par}

\caption{The cubic unit cell ABCDEFGH in a three dimensional nematic. The disclination
ring cuts the upper face ABCD at X and X1. The triangular portions ABC and ADC
must enclose a 1/2 -defect each. }
\end{figure}

{\large Acknowledgment :}{\large \par}

{\large The authors acknowledge a UGC grant F.10 - 17/2001(SR-1) which enabled
us to upgrade the computing facility. }{\large \par}

{\large One of us (S. Dutta) acknowledges financial support from the Council
of Scientific and Industrial Research (}\textbf{\large CSIR}{\large ), India.}{\large \par}


\begin{thebibliography}{10}
\bibitem[13]{enakshi1}{\large E. Mondal and S.K. Roy , Phys. Lett.} \textbf{\large A324}{\large ,
337 (2004), (It may be noted that the mention of topological 1-charge included
by the elementary square plaquettes in this Letter should be replaced by 1/2-pair
charges enclosed by the two elementary triangular plaquettes that form an elementary
square) }{\large \par}
\end{thebibliography}
\end{document}